\documentstyle[12pt,aasms4]{article}
\lefthead{GWINN ET AL.}
\righthead{PROPER MOTIONS FROM GRAVITATIONAL WAVES}
\begin{document}
\input epsf

\title{Quasar Proper Motions and Low-Frequency Gravitational Waves}
\author{Carl R. Gwinn\altaffilmark{1}, T. Marshall Eubanks\altaffilmark{2}, 
Ted Pyne\altaffilmark{3,4},}
\author{Mark Birkinshaw\altaffilmark{3,5}, and Demetrios N. Matsakis\altaffilmark{2}}

\altaffiltext{1} {Physics Department, University of California, Santa Barbara, California 93106, USA}
\altaffiltext{2} {U.S. Naval Observatory, Washington, D.C. 20392, USA}
\altaffiltext{3} {Harvard-Smithsonian Center for Astrophysics, Cambridge, MA 02138, USA}
\altaffiltext{4} {Present address:25 Hammond St. \#1, Cambridge, MA 02138, USA}
\altaffiltext{5} {H.H. Wills Physics Laboratory, Bristol University, Tyndall Avenue, Bristol, BS8 1TL, UK}

\begin{abstract}
We report observational upper limits on the mass-energy of the
cosmological gravitational-wave background, from limits on proper
motions of quasars.  Gravitational waves with periods longer than the
time span of observations produce a simple pattern of apparent proper
motions over the sky, composed primarily of second-order transverse
vector spherical harmonics.  
A fit of such harmonics to measured motions yields a
95\%-confidence limit 
on the mass-energy of gravitational waves 
with  
frequencies $\nu<2\times 10^{-9}$~Hz,
of $<0.11 h^{-2}$ times the
closure density of the universe.
\end{abstract}


\section{INTRODUCTION}

Like variations in refractive index,
the changes in spacetime metric produced by gravitational
waves alter optical path lengths.
Light propagation through gravitational waves preserves sources'
surface brightness and total intensity,
to first order in the wave amplitude (\cite{bon59,zip66,pen66});
but it can produce oscillations in apparent position, at the
period of the gravitational wave (\cite{fak94,dur94,bar94,pyn96,kai96}).
Over time intervals much
shorter then a gravitational wave period, these deflections
cause a characteristic pattern of apparent motions in
the plane of the sky (in astronomical parlance, proper motions)
(\cite{pyn96}).
In this Letter we set stringent upper limits on the energy density of
such waves, using measurements of the proper motions of quasars.

Although detected only indirectly to date, most cosmologists believe
gravitational waves are commonplace.  
Very low-frequency gravitational
waves arise naturally in inflationary cosmologies 
(\cite{rub82,fab83})
and dominate the mass-energy of the universe 
in some of them(\cite{gri93}).
Other possible sources of such a background include
phase transitions in the early universe, networks
of cosmic strings, and collisions of bubbles (\cite{vil81,hog86}).

Various observations set direct or indirect
observational constraints on the spectrum of low-frequency
gravitational radiation.  
These constraints are often expressed in terms of
$\Omega_GW$, the ratio of the mass-energy
density of the gravitational waves to that required to close the universe.
Some observational constraints are sensitive to a narrow range of
frequencies,
and are best expressed in terms of the logarithmic derivative
$d\Omega_{GW}/d\ln\nu$ evaluated at some frequency $\nu$.
Timing of pulsars sets observational
limits for periods as great as
the span of observations (\cite{bac86}).  For the extremely stable millisecond
pulsars the limit corresponds to 
$d\Omega_{GW}/d\ln\nu <10^{-8}$, at frequency $\nu = 4.4\times
10^{-9}$~Hz (\cite{kas94}).  After taking into account
the energy carried away by gravitational waves and the effects of
galactocentric acceleration, orbital periods of binary pulsars are
sensitive to gravitational waves with periods as great as the light
travel time from the pulser.  Current limits from such data sets a
limit of $\Omega_{GW} h^2<0.04$ for $10^{-11}<\nu<4.4\times
10^{-9}$~Hz, and of $\Omega_{GW} h^2<0.5$ for
$10^{-12}<\nu<10^{-11}$~Hz 
(\cite{ber83,tay89,tho96}).
Here, the normalized Hubble
constant is $h=H_0/100~{\rm km~s}^{-1}$.  These measurements are
sensitive to gravitational waves with periods as great as the light
travel time from the pulsar.  
A cosmological
background of gravitational waves at the epoch of recombination can
produce anisotropies of the cosmic background radiation 
(\cite{lin88a,kra92}).  For gravitational-wave spectra typical of
inflation, the anisotropy detected by COBE yields a limit of
$d\Omega_{GW}/d\ln\nu\leq 10^{-11}$, with sensitivity of the
measurement concentrated near $\lambda\approx 2$~Gpc, or $\nu\approx
10^{-17}$~Hz (\cite{bar94}).
Few
constraints 
have been set in the frequecy range $10^{-17}<\nu<10^{-12}$~Hz
Linder \markcite{lin88b}(1988b) 
found that a gravitational-wave background in this frequency
range would affect the galaxy-galaxy correlation function,
but robust calculations would require knowledge
of the ``true'' correlation function,
observable only in the absence of such a background.

In contrast to these techniques, proper motions of extragalactic radio
sources are sensitive to waves of arbitrarily long wavelength, and are
independent of the spectrum and source of gravitational-wave radiation. 
Under the assumption that the
spectrum of gravitational waves is stochastic, the squared proper
motion of distant sources, suitably averaged over the sky, is
proportional to the energy density of the waves,
at frequencies from the inverse of the period of
observation to the Hubble time.

In a previous paper(\cite{pyn96}) we discuss the pattern of
apparent proper motions of distant radio sources, produced by a
gravitational wave.  We use the assumption that the wavelength is
short compared to the distance to the source, and the adiabatic
approximation that the wavelength is less than the Hubble length;
however, because the observed effect is local to the Earth, these
assumptions can probably be relaxed without much change in the
result.
The best present measurements of proper motions 
attain accuracies of a few microarcseconds ($\mu$as) per year 
(\cite{eub96})
corresponding to observational limits on the energy
density of gravitational waves of about that required to close the
universe.  

\section{THEORETICAL BACKGROUND}

Very-long baseline interferometry (VLBI) measures positions of radio sources
by measuring the difference in arrival times 
of their signals at antennas
in different geographical locations (\cite{sha76}).  The
interferometrist assumes that the observations are made in a locally
Minkowski reference frame (allowing for 
the orbital acceleration of the Earth 
and the 
general-relativistic light-bending of the Sun
and planets), and so interprets these observations in the Gaussian
normal reference frame.  The delay $T$ between arrival times measures
the projection of the unit vector pointing toward the source
onto the spacelike baseline vector that connects the antennas.  
Measurement of the delay for many sources on several
baselines allows solution for both source positions and lengths
and orientations of the baselines.

Pyne et al. \markcite{pyn96}(1996) describe the effect of a gravitational wave on a
VLB interferometer: the wave produces variations
in delay $T$, which are interpreted as variations in source position.
A gravitational wave traveling toward $+z$, with the ``$+$''
polarization, produces metric perturbations 
$h \cos(p t) ({\bf \hat x\hat x}- {\bf \hat y\hat y})$
in the background coordinate reference frame,
where $h$ is the dimensionless strain of the wave,
$p$ is its angular frequency, and $t$ is time.  
In the 
interferometrist's Gaussian
normal frame, the observed proper motion ${\vec{\mu}}$ 
of a radio source at position
$(\theta,\phi)$ will be:
\begin{equation}
{\vec{\mu}} = {{hp}\over{2}} \sin(p\eta) 
\sin\theta\left(\cos 2\phi {\bf \hat{{\theta}}} 
- \sin 2\phi {\bf\hat{{\phi}}}\right)
\label{pattern}\end{equation}
Here, $\theta$ measures angle from $+z$,
the direction of propagation of the wave, and
$\phi$ measures azimuthal angle around it, from the $x$-axis; 
the associated unit vectors on the sky are 
${\bf \hat{{\theta}}}$ and 
${\bf \hat{{\phi}}}$.
Proper time in the Gaussian normal frame is $\eta$.
We take $h$ to be real,
and allow the origin of time $\eta$ to express
the phase of the wave.
Fig. 1 shows the pattern of proper motions
that this gravitational wave produces.

The properties of this pattern of proper motions
are not simple under
rotation or superposition. 
However, the transverse vector spherical harmonics 
${\bf Y}^{({\rm E},
{\rm M})}_{\ell,m}$ form an
orthonormal basis for vector fields on a sphere, with well-understood behavior
under rotation and superposition.  
Expanded in such harmonics,
Eq.\ (\ref{pattern}) takes the form:
\begin{eqnarray}
{\vec{\mu}} = ph &&\sin(p\eta)\nonumber\\
\times\Bigl\{ && +{{\sqrt{5\pi}}\over{6}}
\bigl[{\bf Y}^{\rm E}_{2,+2}+{\bf Y}^{\rm E}_{2,-2} 
-{\bf Y}^{\rm M}_{2,+2}+{\bf Y}^{\rm M}_{2,-2}\bigr]\\ \label{tvshexp}
&&-{{\sqrt{70\pi}}\over{60}}
\bigl[{\bf Y}^{\rm E}_{3,+2}+{\bf Y}^{\rm E}_{3,-2}
-{\bf Y}^{\rm M}_{3,+2}+{\bf Y}^{\rm M}_{3,-2}\bigr]+...\Bigr\}.\nonumber
\end{eqnarray}
Here we use convention of Mathews \markcite{mat62,mat81}(1962,1981)
for transverse vector spherical
harmonics.
These fall into 2 categories,
with one family, commonly denoted ``poloidal'', ``potential'', or ``electric''
pointing down the gradients of scalar spherical harmonics;
and the other, known as ``toroidal'', ``stream'', or ``magnetic''
pointing perpendicular to their gradients.
We denote these categories as
``E'' and ``M'', respectively
(\cite{mat62,mat81}), with the 
notation $({\rm E},{\rm M})$ meaning ``E or M''.
Note that our ${\bf Y}^{\rm M}_{\ell,m}$
is the ${\bf X}_{\ell,m}$ of Jackson \markcite{jac76}(1975).
The
normalization condition is:
\begin{eqnarray}
4\pi \langle {\bf Y}^{{\rm E}}_{\ell,m} {\bf Y}^{{\rm E}*}_{i,j}\rangle = &&
4\pi \langle {\bf Y}^{{\rm M}}_{\ell,m} {\bf Y}^{{\rm M}*}_{i,j}\rangle = 
\delta_{\ell i} \delta_{m j}\\ \label{normcond}
\langle {\bf Y}^{{\rm E}}_{\ell,m} {\bf Y}^{{\rm M}*}_{i,j}
\rangle = &&0.\nonumber
\end{eqnarray}
Here $\langle...\rangle$ denotes an average over the sky.

Because the ${\bf Y}^{({\rm E},{\rm M})}_{\ell,m}$ form an orthonormal
basis for vector fields on a sphere,
they can represent 
proper motions
due to a single wave, or
any spectrum of waves.
We use the expansion
\begin{equation}
{\vec{\mu}} = {\sum_{\ell,m}} 
\Bigl[a^{\rm E}_{\ell,m} {\bf Y}^{\rm E}_{\ell,m} 
+ a^{\rm M}_{\ell,m} {\bf Y}^{\rm M}_{\ell,m}\Bigr]
\label{defas}\end{equation}
to define the 
coefficients $a^{\rm E}_{\ell,m}$ and $a^{\rm M}_{\ell,m}$
in terms of the proper motion ${\vec{\mu}}$,
observed over a small fraction of a gravitational wave period.
The $a^{\rm (E,M)}_{\ell,m}$ must satisfy
(\cite{mat81}):
\begin{equation}
a^{\rm E}_{\ell,+m}=(-1)^m a^{\rm E*}_{\ell,-m}\quad
a^{\rm M}_{\ell,+m}=(-1)^{m+1} a^{\rm M*}_{\ell,-m}
\label{realcond}\end{equation}
because ${\vec{\mu}}$ is real.

The squared proper motion 
averaged over the sky
is related to the energy density of the wave
$T_{GW}$ by
\begin{equation}
\langle\mu^2\rangle ={{1}\over{12}}p^2 h^2
={{8\pi}\over{3}}{{G}\over{c^2}}T_{GW},
\label{nutgw}\end{equation}
Here $\langle..\rangle$ includes an average over many wave
periods, as well as over the sky.
Expansion of the mean square proper motion in vector spherical 
harmonics shows that 5/6 of the mean square motion resides 
in the $\ell=2$ harmonics:
\begin{equation}
\frac{1}{4\pi}\sum_{m} 
\left(|a^{{\rm E}}_{2,m}|^2 + |a^{{\rm M}}_{2,m}|^2\right)
= {{5}\over{6}} \langle\mu^2\rangle.
\label{562nd}\end{equation}
Because the summed, squared moduli of the $a^{\rm (E,M)}_{\ell,m}$
remain constant under rotation for for each $\ell$, and
for ${\rm E}$ and ${\rm M}$ harmonics
separately, Eq.\ (\ref{562nd}) holds for a gravitational wave propagating in
any direction with any polarization.

If the phases, amplitudes and
directions of the many superposed waves in the spectrum
are random at
the Earth, as is to be expected for a cosmological background of
gravitational waves, the random-phase approximation shows that
Eqs.\ (\ref{nutgw}) and (\ref{562nd}) will hold statistically,
if $p^2h^2$ is replaced by
the integral $\int p^2 h(p)^2 d^3p$,
and if the average of the squared proper motion
over time and sky is replaced by an average over
an ensemble of gravitational-wave spectra 
(e.g. \cite{goo85}).
Here $h(p)^2$ is the spectral density of
the square of the dimensionless strain.
Thus, Eq.\ (\ref{562nd}) shows that 5/6
of the mean square proper motion due to the low-frequency
gravitational wave background appears in second-order transverse
spherical harmonics, and 
${\rm E}$ and ${\rm M}$ harmonics contribute equally.
We can express the energy density in such a stochastic background
of gravitational waves as $\Omega_{GW}$, its ratio to the
closure density
of the universe, $T_{CL}=3 c^2 H_0^2/8\pi G$;
and we can relate 
$\Omega_{GW}$ to proper motion in second-order transverse harmonics: 
\begin{equation}
\Omega_{GW}={{1}\over{H_0^2}}\langle\mu^2\rangle
={{6}\over{5}}{{1}\over{4\pi}}{{1}\over{H_0^2}}\sum_{{\rm(E,M)} m} 
|a^{{\rm (E,M)}}_{2,m}|^2
\label{omegaa2}\end{equation}
We seek to measure the $a^{\rm (E,M)}_{2,m}$
and to thus measure, or set limits on,
$\Omega_{GW}$.

\section{OBSERVATIONS AND DATA REDUCTION}

We searched for the pattern expected for gravitational waves in proper motions 
of extragalactic
radio sources.  We estimated second-order spherical harmonics from a
proper-motion solution using the US
Naval Observatory comprehensive database of VLBI
observations. This contains most of the suitable astrometric and geodetic
observations of extragalactic radio sources made with the Mark III and
compatible VLBI systems\cite{cla85}.
Eubanks and Matsakis \cite{eub96}
describe the data analysis
in detail; here
we present a brief overview.

The database comprises 1,469,793 observations of delay $T$ and its
time derivative $\dot T$ with different baselines and sources.  The
observations were made between 3 August 1979 and 13 February 1996,
with more than 70\% after 1990.  Proper motions were determined for
the 499 sources observed at
more than 1 epoch.  
In fitting for the coefficients of transverse spherical
harmonics we used the 323 sources known to be extragalactic
by their measured redshifts.
Their mean redshift is 1.22.

Reduction of this data to source
positions and proper motions follows standard procedures for
astrometric and geodetic VLBI data (e.g. \cite{cla85}).
We removed the best models for earth orientation, tidal deformation,
and location within the solar system; atmospheric propagation; and
solar-system effects of general and special relativity; before fitting
for proper motion.  
Simultaneous observations at 2 frequencies allowed calibration of 
the propagation delay through the ionosphere. The
solutions for proper motions also included $\approx 400,000$
``nuisance'' parameters, including those describing positions of
sources and stations; clock behavior at different stations;
optical path length though the atmosphere in a vertical direction and
its horizontal gradients; unmodeled motions of the Earth, including
variations in rotation, polar motion, nutation, and precession; and
gravitational deflection of light by the Sun \cite{eub96}. 
These nuisance parameters are chosen to have little covariance
with source and station positions,
and thus to proper motions.
After the fit, the weighted root-mean-square of the scatters
of the residuals in delay and in delay rate were 31 ps and
93~fs~s$^{-1}$, respectively.

Effects of source structure and variations in atmospheric path length
are among the expected systematic errors.
The sources are highly energetic neighborhoods of active galactic
nuclei.
The well-known motions of 
jets from such
sources can mimic proper motion, particularly in cases where the
stationary core component is weak or absent 
(e.g. \cite{gui95}).  
Such motions are expected to be many 
$\mu{\rm as~yr}^{-1}$, but should not be correlated over the sky,
so that they increase noise in the measurements but do not bias
lower-order transverse spherical harmonics.
Atmospheric path length is removed via models incorporating
contemporaneous meteorological data, and the residual path length 
and its horizontal gradients are
parametrized and estimated directly from the VLBI solution.  Remaining
atmospheric delay errors might show angular
dependences that could mimic those
expected for gravitational waves.

We then fit 
transverse spherical harmonics to the measured proper motions.
The fit minimized $\chi^2$,
the summed, squared differences of
proper motions and the motions modeled by 
Eqs.\ (\ref{defas}) and (\ref{realcond}).
These differences were weighted by the expected
errors, given by the standard errors from the fit to delay and delay
rate, scaled by a factor of 1.35, and with an assumed error added in
quadrature (in each component of proper motion) of 
30~$\mu{\rm as}~{\rm yr}^{-1}$.
This reweighting serves to make the $\chi^2$ per
degree of freedom about 1,
as expected; it accounts, in particular,
for effects of variations in source structure.
Comparison of fits to subsets of the data suggested
this scaling and addition in quadrature.
Fits to subsets of the data did not change
the fitted values significantly, 
although the standard errors increased.

The fit included coefficients for $\ell=1$ transverse spherical
harmonics, as well as the $\ell=2$ harmonics characteristic of gravitational
waves. The ${\rm M}$ $\ell=1$ harmonics correspond 
to a rotation, of no physical significance, 
as it is not separable from
the Earth's rotation; 
the ${\rm E}$ $\ell=1$ harmonics correspond to acceleration
not included in models used for data reduction.
This acceleration is an interesting cosmological
parameter: it is sensitive to the galactocentric acceleration of the
Solar System, as well as any acceleration of the Milky Way relative to
distant quasars.

The measured values of the coefficients of second-order transverse spherical
harmonics are statistically indistinguishable from noise.
The residual $\chi^2$ to the fit was 717,
distributed among 627 degrees of freedom.
Table 1 summarizes results of the fit,
and Fig. 1 shows the results for the $\ell=2$ harmonics
in graphical form.
The fitted coefficients 
are combinations of $a^{\rm (E,M)}_{\ell,m}$
that satisfy Eq.\ (\ref{realcond}) to produce real motions.
The mild migration of sources away from the
ecliptic in Fig. 1 may reflect
effects of the Sun on atmospheric propagation.

\section{RESULTS}

A particular value of $\Omega_{GW}$
corresponds to an ensemble of possible 
gravitational-wave spectra and coefficients $a^{\rm (E,M)}_{2,m}$.
Eq.\ (\ref{omegaa2}) relates $\Omega_{GW}$ to the mean over the ensemble
of the summed, squared coefficients.
The measurement errors given in the last column of
Table 1 also contribute to the measured $a^{\rm (E,M)}_{2,m}$.
We take these contributions into account to find the 
value of $\Omega_{GW}$ that yields
summed, squared coefficients less than those we observe 
for only 5\% of gravitational spectra,
and adopt this as our upper limit.
This procedure includes effects of sample or cosmic
variance.  With 95\% confidence, we find that
$\Omega_{GW}< 0.11~h^{-2}$, where $H_0=h 100~{\rm km~s}^{-1}$. 
This limit holds
for a stochastic spectrum of gravitational waves integrated over all
frequencies less than half the inverse of the 10-yr span of the
observations, or about $2\times 10^{-9}$~Hz.

Considerable improvement in this limit 
should be possible over the next decade. The 
measurement accuracy for 
proper motions is proportional to the
duration of observations to the 3/2 power. Because 
most data were acquired in
the last 5 to 10 years, an additional
decade of observing should improve the bound on $\Omega_{GW}$ by a
factor of 3 to 8. 
Moreover, we can reduce effects of source structure
by choosing additional sources with little 
structure, from among the thousands observable astrometrically 
with the Very Long Baseline Array; and by using models
derived from images of the sources to correct for effects of
structure directly. Together these improvements may be expected to
reduce the formal errors of measurements of $\Omega_{GW}$ by a factor
of 10 to 100, within the next 10 years.

\acknowledgements
We thank J. Hartle for useful suggestions.
This work was supported in part by the National Science Foundation
(AST90-05038 and AST92-17784).

\newpage
\begin{figure}[t]
\small
\figurenum{1}
\plotone{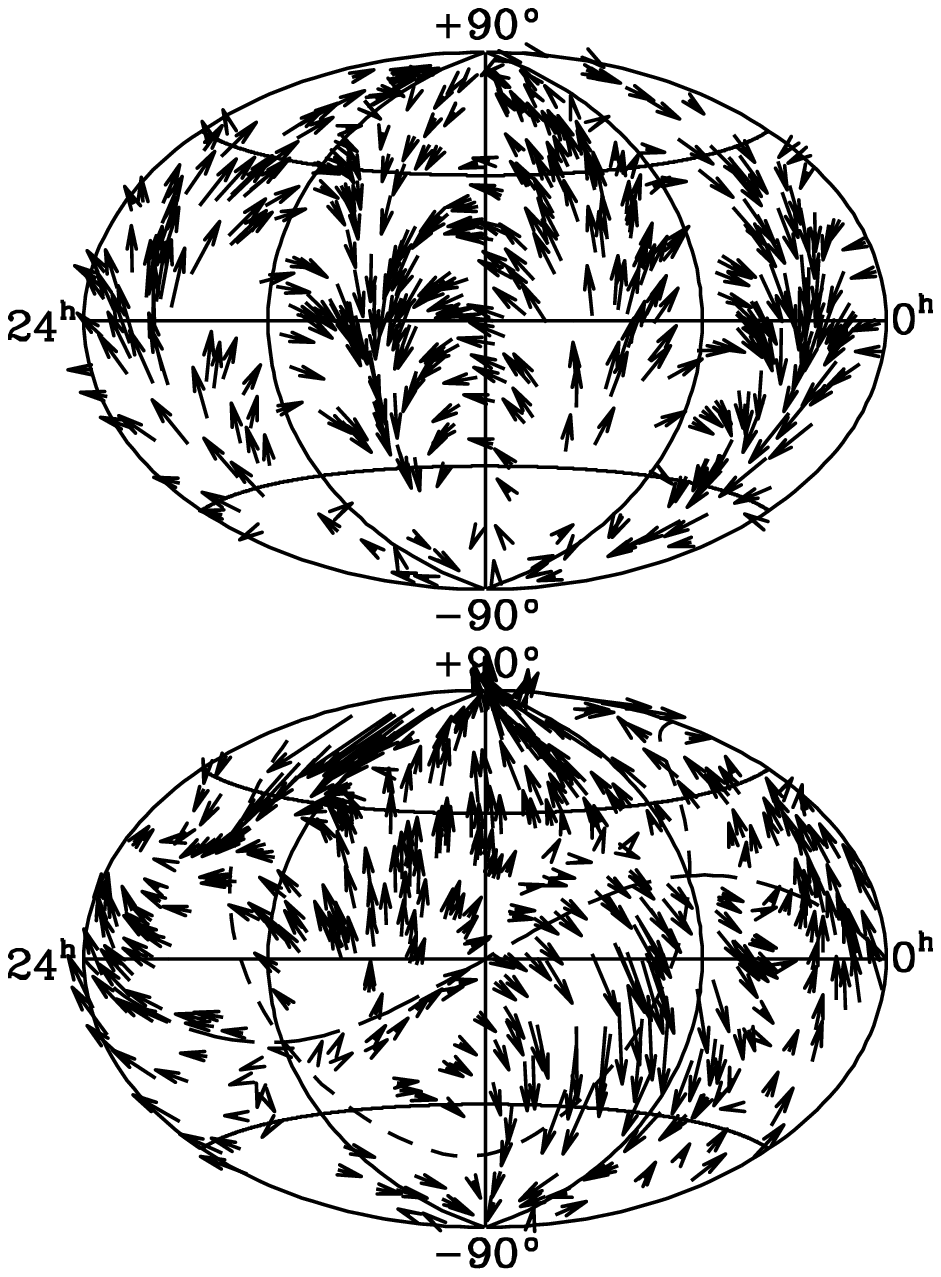}
\figcaption[2vert.ps]{
Upper: Proper motions expected for a single gravitational wave.
The metric perturbation is
$h \cos(p(cz-t)) ({\bf \hat x\hat x}- {\bf \hat y\hat y})$, 
with ${\bf \hat z}$ toward declination $90^{\circ}$ and 
${\bf \hat x}$ toward right ascension $0^{\rm h}$.
Lower: Fitted coefficients of the second-order ($\ell=2$) transverse
spherical harmonics, displayed as proper motions
at locations of sources with
measured proper motions. Arrow lengths in degrees equal proper
motion in $\mu{\rm as}~yr^{-1}$.
Coefficients are not shown for $\ell\neq 2$. Curves
show the ecliptic (long dashes) and galactic plane (short
dashes). 
The fitted coefficients are not statistically significant, so that
the observed pattern of motions is consistent with filtered noise.
\label{2vert}}
\end{figure}

\newpage
\begin{deluxetable}{l r r}
\small
\tablecolumns{3}
\tablewidth{400pt}
\tablecaption{Observed Coefficients of Transverse Harmonics \label{fitted}}
\tablehead{
\colhead{}         &\colhead{Fitted}          &\colhead{Standard}\\
\colhead{Parameter}&\colhead{Value}           &\colhead{Error}\\
\colhead{}         &\colhead{($\mu{\rm as~yr}^{-1}$)}&\colhead{($\mu{\rm as~yr}^{-1}$)}\\
}
\startdata
\multicolumn{3}{l}{Acceleration\tablenotemark{a}}\nl
$\ddot x$ & 1.9& 6.1 \nl
$\ddot y$ & 5.4& 6.2 \nl
$\ddot z$ & 7.5& 5.6 \nl
\tablevspace{12pt}
\multicolumn{3}{l}{Second-Order Transverse Vector Spherical Harmonics\tablenotemark{b}}\nl
$	{a^{\rm E}}_{2,0}                                      $& 12.1& 16.6\nl
${{{1}\over{\sqrt{2}}}} ({a^{\rm E}}_{2,+1}-{a^{\rm E}}_{2,-1})$&-13.6& 14.9\nl
${{{i}\over{\sqrt{2}}}} ({a^{\rm E}}_{2,+1}+{a^{\rm E}}_{2,-1})$&-21.7& 15.6\nl
${{{1}\over{\sqrt{2}}}} ({a^{\rm E}}_{2,+2}+{a^{\rm E}}_{2,-2})$&  4.2& 12.5\nl
${{{i}\over{\sqrt{2}}}} ({a^{\rm E}}_{2,+2}-{a^{\rm E}}_{2,-2})$& -0.6& 13.3\nl
$    i  {a^{\rm M}}_{2,0}                                      $&  1.7& 14.4\nl
${{{i}\over{\sqrt{2}}}} ({a^{\rm M}}_{2,+1}-{a^{\rm M}}_{2,-1})$& 17.8& 15.0\nl
${{{1}\over{\sqrt{2}}}} ({a^{\rm M}}_{2,+1}+{a^{\rm M}}_{2,-1})$&-28.2& 15.3\nl
${{{i}\over{\sqrt{2}}}} ({a^{\rm M}}_{2,+2}+{a^{\rm M}}_{2,-2})$&-17.6& 13.9\nl
${{{1}\over{\sqrt{2}}}} ({a^{\rm M}}_{2,+2}-{a^{\rm M}}_{2,-2})$& 10.7& 15.0\nl
\enddata
\tablenotetext{a}{Effects of acceleration of the solar system 
barycenter relative
to the observed extragalactic radio sources corresponds to first-order ($\ell=1$)
harmonics.
Rotations 
also correspond to first-order harmonics;
our fits included these, but we do not report them 
as they cannot be separated from
the Earth's rotation.}
\tablenotetext{b}{Second-order ($\ell=2$) harmonics correspond to 
effects of low-frequency gravitational waves.
Otherwise identical coefficients $a^{\rm (E,M)}_{\ell,m}$ with opposite
sign of $m$ have been combined to reflect
the fact that the fitted motions are purely real.}
\label{table1}
\end{deluxetable}
\end{document}